# Electroconvective flow in presence of polyethylene glycol oligomer additives

Arpita Sharma,[a] Ankush Mukherjee,[b] Alexander Warren,[a] Shuo Jin,[a] Gaojin Li,[a] Donald L. Koch,[a] & Lynden A. Archer*[a]

[a]*School of Chemical and Biomolecular Engineering, Cornell University, Ithaca, NY 14853*
[b]*School of Mechanical and Aerospace Engineering, Cornell University, Ithaca, NY 14853*
*Correspondence Email\*: laa25@cornell.edu*

## Abstract

Metal electrodeposition in batteries is fundamentally unstable and affected by different instabilities depending on operating conditions and chemical composition. Particularly at high charging rates, a hydrodynamic instability called electroconvection sets in that aggravates the situation by creating non-uniform ion flux and preferential deposition at the electrode. Here, we experimentally investigate how oligomer additives interact with the hydrodynamic instability at a cation selective interface. From electrochemical measurements and direct visualization experiments, we find that electroconvection is delayed and suppressed at all voltage in the presence of oligomers. Our results also reveal that it is important to consider the role of polymers at the interface, in addition to their bulk effects, to understand the stabilization effect and its mechanism.

1. **Introduction**

Electroconvection is a hydrodynamic instability that plays an important role in several processes involving charge transfer across an ion selective interface.[1] These include ion separation in electrodialysis, electrodeposition in high energy density batteries, metal electroplating and desalination for water purification. According to current understanding, the origin of electroconvection is attributed to formation of an extended space charge layer, which interacts with the imposed electric field to create a body force that drives complex fluid motion near interfaces.[2] This additional mode of mass transfer allows the ion flux to overcome classical limitations imposed by concentration polarization and manifests as an higher-than-expected ion flux to the interface that drives a so-called over-limiting current. Experimentally, electroconvection phenomena has been observed by recording current-voltage (*i*-V) profiles in electrochemical cells bounded by ion-selective interfaces at electrodes and cation/anion exchange membranes.[3] The resultant *i*-V curve typically manifests 3 distinct regions which correspond to: ohmic (under limiting) ion transport, where current is proportional to voltage and bulk electroneutrality is maintained; a limiting region, where current is limited by the rate of ion diffusion, and an ion depleted space charge region starts to form at the interface; and lastly, an over-limiting region, where current overcomes the diffusion limitation by formation of convective rolls near the interface. Direct numerical simulation studies by Mani and coworkers have revealed the structure and dynamics of these electroconvective vortices and demonstrated that they undergo a transition from a steady to unsteady convection at high voltages where the flow becomes multiscale and chaotic.[4] Density gradients associated with ion separation can further complicate the flow pattern and analysis, especially in solutions with high Rayleigh numbers.[5] Karatay et al. reported a strong coupling between gravity and electroconvection that depends on their relative direction and can expediate and destabilize electroconvection by the onset of Rayleigh- Bernard convection.[6]

Electroconvective flow is also reported to be strongly influenced by the electrical, geometrical, and chemical properties of the system.[1] For instance, the thickness of the extended space charge layer, which is the driving force for electroconvection, is affected by salt concentration, applied voltage and cell geometry as $\tilde{\delta}_s \approx L^{\frac{1}{3}}(\tilde{V}\tilde{\delta})^{\frac{2}{3}}$ where $\tilde{\delta}_s$ is the extended space charge layer thickness, $\tilde{V}$ is the applied potential, $L$ is the interelectrode spacing and $\tilde{\delta}$ is Debye screening length. Theoretical studies by a number of groups have shown that by varying these parameters it is possible to achieve a stronger and more complex flow at higher $\tilde{\delta}_s$ values.[4,7,8] These findings have been verified by direct visualization studies where tracer particles and PIV (particle image velocimetry) analysis reveal that micro-scale flow characteristics like vortex size, growth rate and velocity distribution can be profoundly altered through simple changes in experimental conditions, including physical properties of an electrolyte and voltages employed.[9,10]

An already significant body of studies have discussed strategies for modulating the electroconvective flow - to suppress or enhance it, depending on specific applications. In electrodialysis, electroconvection is considered advantageous as it provides an additional mode of mass transfer and promotes ion mixing. Several techniques including the use of patterned membranes or an external shear velocity have been proposed to further enhance the flow.[11,12,13] Unlike electrodialysis, applications requiring controlled electrodeposition of metals at high rates, including for the manufacture of thin metal electrodes or for sustaining long-duration charge/discharge reversibility in batteries at high and intermediate rates (relative to the diffusion limited current density), electroconvection is undesirable and can adversely affect control performance by creating non-uniform ion flux at the interface. A particular concern in the latter situation is that coupling between electroconvection and an evolving, complex substrate morphology enhances instability, leading to rapid loss of uniformity and enhanced out-of-plane growth of the deposit.[14,15] In battery cells the non-uniform deposits are dangerous because they may grow out to bridge the inter-electrode space, short circuiting the cell. Strategies for eliminating the obvious safety and shortened lifetimes of such cells are discussed in a number of excellent recent reviews.[16–18] As an example, to overcome the compound challenge associated with deposition in situations where electroconvection and morphological instabilities are coupled, gel membranes which prevent electroconvection by immobilizing the depletion diffusion layer or by using solid electrolytes made of cross-linked polymer chains have been reported.[19,20] Liquid electrolytes infused with small amount of ultrahigh molecular weight polymers like poly (methyl methacrylate) ($M_w = 0.996 \times 10^6 g/mol$) and polyethylene oxide ($M_w = 8 \times 10^6 g/mol$) have also been reported to be effective in suppressing electroconvection and improving deposition morphology in several metals.[14,21,22] It was found that polymer entanglements played a key role in enhancing system stability. These entanglements are therefore thought to form a mesh like network in the electrolyte which suppress the electroconvective flow, which in turn increases the critical voltage for onset of unstable deposition. As intriguing, however, are very recent reports which show that very low molecular weights polymers, specifically polymers with molecular weight ($M_w$) well below the entanglement molecular weight ($M_w < M_e$), can be effective in promoting long-term cycling of battery cells that use metals as anodes.

In this work, we address an open question related to why/how such low molecular weight polymers are effective in suppressing electroconvection, and on that base identify opportunities as well as limits for using the approach for extending battery lifetime. Here we focus on liquid electrolytes infused with a low molar mass polyethylene glycol ($M_w = 300 \ g/mol$) additive. The chosen molecular weight is well below the entanglement molecular weight (2000 g/mol) of the undiluted PEG polymer which, means that irrespective of polymer concentration the electrolyte solutions employed in our study are unentangled. We investigate the electrochemical response, surface adsorption dynamics, flow dynamics and physical properties of the electrolytes to understand the role of polymer at the interface and bulk on electroconvection. We also compare our experimental findings with results from theoretical analysis to understand the fundamental aspects of the stabilization mechanism.

2. **Experimental Methods**

Zinc sulfate and Polyethylene glycol ($M_w = 300 g/mol$, denoted PEG300 henceforth) were purchased from Sigma Aldrich. Zinc foil (thickness 0.25mm) used for electrodes was purchased from Alfa Aesar. To decouple the hydrodynamic (electroconvection) and morphological instabilities, electrochemical cells containing an ion-selective interlayer membrane (Nafion® purchased from Sigma Aldrich). Depending on the experiment, either free standing Nafion® NRE-212 (thickness 0.05mm) or a thin film (thickness ~ 2µm) composed of Nafion™ perfluorinated resin solution spin-coated on the working electrode was used. Current voltage curves were obtained using a 3-electrode Rotating Disk Electrode chamber (Pine Research) with Nafion® coated glassy carbon as the working electrode, zinc foil as the counter electrode and Ag/AgCl as the reference electrode. No rotation was applied to the 3-electrode setup. A scan rate of 1mV/s was applied using CH Instruments Electrochemical Analyzer Potentiostat.

Visualization experiments were conducted in a lab made cell depicted in **Fig. 3c** using Zeiss Elyra Super Resolution microscope. To minimize 3D flow, the gap between the bottom glass slide and top coverslip of the visualization cell was fixed at 250 µm.[23] The electrode size was approximately 2 cm and the spacing between the electrodes was fixed at 5mm. The working zinc electrode in the cell was covered with a Nafion® membrane, which served as an ion-selective interface. The entire setup was sealed using a Press-To-Seal silicone isolator from Millipore Sigma. In this configuration, the liquid electrolyte was injected into the sealed cell using a syringe. FluoSpheres™ Carboxylate-Modified Microspheres (diameter = 500 nm) (Thermo Fischer Scientific) were used as tracer particles to visualize the velocity field in the electrolyte. The tracer particle charge was deduced to be -25.6mV, from independent measurements of the zeta potential using a Nano ZS90 Zetasizer (Malvern Panalytical). During visualization experiments, the cell was placed horizontally to limit complications from the Rayleigh-Bernard type convection; in this orientation gravity is perpendicular to the electric field.[5,6] A constant voltage was applied across the visualization cell with a BK Precision power supply and videos were recorded at a high frame rate of 25fps under 5X magnification. The videos were recorded for approximately 5 minutes after applying the potential. The individual frames from the videos were extracted and processed through a particle tracking package (PIVlab) in Matlab. The PIVlab software uses a cross-correlation algorithm over multiple passes and finally provides velocity vectors for every 15 µm X 15 µm section of the frame. Outliers were removed and then replaced via interpolation. RMS velocity variation with distance from Nafion$^{TM}$ surface was calculated by averaging in both y coordinate and time, using the formula: $<u> = (\frac{1}{y_1-y_0}\frac{1}{t_1-t_0}\int_{y_0}^{y_1}\int_{t_0}^{t_1} u^2 dt dy)^{1/2}$. A similar PIV analysis has been used in previous studies to extract velocity vectors to characterize electroconvective flow structure.[24,25]

As discussed in the introduction, the oligomers have a large effect on both the onset condition and intensity of electroconvection that appear incommensurate with their relatively small impacts on the bulk electrolyte viscosity. We hypothesized that their influence was primarily interfacial, and a result of a concentrated polymer coating formed by spontaneous adsorption at the electrode or ion-selective interface. Quartz Crystal Microbalance (QCM-D) measurements were performed with Q-sense Explorer system (Biolin Scientific AB) at 25°C to study this effect in detail. Fundamental frequency and three harmonics (third, fifth and seventh) were recorded simultaneously along with the corresponding dissipation factors. Gold coated AT-cut quartz crystal (fundamental frequency ~ 4.95 MHz) was purchased from Q-sense (QX301). 1mM $ZnSO_4$ was used as the buffer solution and a stable baseline was obtained before switching to different PEG300 concentration electrolytes. The data obtained from QCM-D was modeled using the Q-tools software (Q-sense) using descending incremental fitting. 'Voigt' viscoelastic model was used to model 3$^{rd}$, 5$^{th}$ and 7$^{th}$ harmonics.[26,27] Baseline fluid density and viscosity were taken to be 1000 kg/m$^3$ and 0.001 Pa-s, respectively. The inbuilt Voigt Model requires a user defined layer density and for this case adsorbed layer density was assumed to be 1125 kg/m$^3$ (which is the density of the PEG300) for calculating viscoelastic properties. In addition to obtaining the layer properties from the inbuilt Voigt model in Q-tools software, we also directly solve the Voigt model equations in Matlab to get adsorbed layer properties. The equations, details and results of which are provided in supporting information (Table S4).[27] Galvanostatic Electrochemical Impedance Spectroscopy (GEIS) experiments were conducted using SP-50e Potentiostat from Biologic. Symmetric Zinc coin cells (electrode area = 1.27 cm$^2$) with Whatman® glass microfiber separator were assembled and a alternating current of 1 µA amplitude was applied. The impedance data was recorded in 1MHz-100mHz frequency range and was modeled using BioLogic EC-Lab® Software. Modeling details are provided in supporting information.

3. Results

3.1 Electrochemical Response

A key result of our study is reported in **Fig. 1**, which shows that the oligomer concentration has a large effect on the current voltage response measured in our three-electrode cell. Specifically, we observe 3 primary effects. Firstly, oligomers delay the onset of overlimiting conduction, as seen by the increase in the width of the diffusion limited plateau ($\Delta$V) observed in the *i*-V curves. Detailed discussion on measuring limited plateau ($\Delta$V) is provided in **Fig. S1**. Secondly, from **Fig. 1b**, we observe that oligomers suppress the over-limiting current, as indicated by a decrease in the normalized current at different voltages ( $i_N = \frac{i}{i_L}$; $i_N = 1$ indicates no electroconvection). Finally, we observe that this delayed onset and suppression of overlimiting current increases in a nonmonotonic manner with oligomer concentration — it at first increases nonlinearly with oligomer concentration but interestingly, saturates at higher concentrations. We analyze the latter effect in more detail in **Fig. 1c**, which reports the measured voltage window ($\Delta V$) vs bulk viscosity ($\eta$) for different polymer concentrations. The figure shows that the voltage window increases as $\Delta V \sim \eta^{1.8}$ initially but changes to $\Delta V \sim \eta^0$ at oligomer concentrations above 20%. It is also noteworthy that the initial scaling, $\Delta V \sim \eta^{1.8}$, is different from $\Delta V \sim \eta^{0.5}$ scaling reported in previous experimental studies with high molecular weight polymer additives and in theoretical studies that considered simple Newtonian liquid electrolytes.[14,22,28] This finding is significant because the $\Delta V \sim \eta^{0.5}$ is achieved in any electrolyte in which the bulk viscosity is increased while other properties remain the same. The much stronger scaling observed in our experiments implies that other physics are important in the present case. We also observe a similar result for a higher salt concentration (10mM ZnSO$_4$) with PEG300 additive with and without the Nafion membrane (**Fig. S2** and **S3**) which indicates the need for further investigation to fully understand the role of oligomer additives. Previous experimental studies of polymer additives in liquid electrolytes have reported an increase in interfacial/charge transfer resistance to be an indicator of polymer adsorption on the surface. To investigate this possibility, we conducted galvanostatic impedance experiments (GEIS) in symmetric zinc coin cells. The resulting Nyquist impedance plots for different PEG300 concentrations are reported in **Fig. S5**. An equivalent circuit model ($R1 + R2/Q2 + R3/Q3$) was used to fit the data and the extracted charge transfer resistance is reported in **Fig. 1d**. The results show a significant increase in charge transfer resistance, consistent with formation of a polymer layer at the interface.[29]

3.2 Quartz Crystal Microbalance with Dissipation (QCM-D)

Quartz Crystal Microbalance studies were used to interrogate the physical characteristics and properties of the adsorbed layer. In particular, QCM-D was employed as a sensitive microbalance capable of detecting adsorbed mass and, through modeling, extract layer properties by measuring shifts in resonant frequency of a piezoelectric quartz sensor.[30] **Fig. 2b** shows the raw QCM-D frequency and dissipation shift for 1mM ZnSO$_4$ + 20% PEG300 electrolyte. A negative shift in frequency is observed, indicating mass addition to the sensor. We also find a high dissipation value and a significant difference in the frequency shift for different harmonics which indicates that the adsorbed layer is viscoelastic in nature (instead of a rigid mass addition). Based on these observations, we use the inbuild single layer 'Voigt' model to estimate the layer properties.[27] The model assumes a homogeneous layer formation and complete coverage of the sensor. While these assumptions may not be accurate for real systems, the Voigt model has been widely used to understand viscoelastic layer properties in QCM-D studies.[31–34] A good fit is determined by low error between experimental and fitted data and the results for 3$^{rd}$,5$^{th}$ and 7$^{th}$ harmonics for 1mM ZnSO$_4$ + 20% PEG300 are reported in **Fig. S6**. **Fig. 2c** shows the modeled layer viscosity and thickness as a function of polymer concentration. We find that the layer viscosity is close to the bulk electrolyte viscosity as measured by rheological experiments (Table S1). The layer thickness estimated from the inbuilt Voigt model varies between 100- 900 nm depending on PEG300 concentration. A larger layer thickness indicates more polymer near the interface but because of the underlying assumptions of Voigt model (a homogenous layer with fixed density, 1125 kg/m$^3$ in this case) the actual adsorbed layer thickness cannot be accurately measured by QCM-D. A comparison of Voigt layer thickness with a dry layer thickness calculated from Ellipsometry is reported in Table S2.

In addition to plotting the direct outl from the inbuilt Voigt model in Q-tools software, we also directly solve the Voigt model equations in Matlab.[27] As previously discussed, the Q-tools software requires the user to enter the value of either density or thickness. It also assumes that the properties of the bulk medium remain completely unchanged when oligomer electrolyte is flown in the QCM-D chamber. To overcome some of these assumptions, we directly solve the QCM-D equations without these inputs and assumptions using Matlab. Here, we took the changes in frequency and dissipation for the 3$^{rd}$, 5$^{th}$ and 7$^{th}$ harmonics along with the bulk density and viscosity as inputs. The bulk density and viscosity are assumed to vary on a linear scale based on polymer concentration. These linear functions are obtained from the properties of base electrolyte (without PEG) and the PEG300 additive electrolyte, with known volume fraction and properties as measured in experiments. We solve for the layer volume fraction, layer elasticity and layer height and the results for which are shown in **Fig. 2b** and Table S4. We find that our QCM model values for layer viscosity and thickness are in the same order of magnitude as the inbuilt Voigt model results. However, we found that the elasticity of the layer is negligible in all cases as reported in Table S4. This indicates that the polymer forms a thin layer with higher volume fraction and viscosity adjacent to the sensor. Furthermore, a large layer thickness and the low value of elasticity suggests that the polymer remains dissolved in solution but with an enhanced concentration relative to the bulk. This observation is crucial as it provides an important design principle for the theoretical model and possible stability mechanism discussed in the later section.

3.3 Visualization Studies

To investigate how the PEG300 oligomer additive impacts the evolution of electroconvective flow in liquid electrolytes, we conducted visualization experiments at different applied potentials. Three sample videos spanning different voltages and concentrations are provided with the supporting data. The key results of visualization experiments are reported in **Fig. 3**. Firstly, we find that the electroconvection develops progressively more slowly in electrolytes containing the oligomer additives and the apparent onset time is delayed markedly, particularly at low applied voltages (see **Fig. 3a**). For instance, no flow was observed for 1mM ZnSO$_4$+30% PEG300 electrolyte at 1V for up to 1 hour following imposition of the electric field. In contrast, in electrolytes that do not contain oligomer a noticeable flow pattern is apparent in as little as 1 minute after applying the potential! The delay time, however, reduces significantly at higher voltages.

The results reported in the inset to **Fig. 3a** report that the onset time is a strong decreasing function of voltage, $t_{onset} \sim V^{-1.88}$; here $t_{onset}$ is the time at which the first sign of flow/particle motion is detected near the Nafion® interface. The scaling increases slightly to $t_{onset} \sim V^{-2.1}$ for the 1mM ZnSO$_4$+30% PEG300 case. This nearly quadratic scaling is qualitatively consistent with the fact that the applied voltage is the driving force for electroconvection. Two factors may contribute to the onset time. First, a space charge layer must develop near the electrode to provide the electro-osmotic force driving electroconvection. Subsequently, perturbations to the fluid flow and ion and oligomer concentrations must develop. We consider the development of the space charge layer here and will discuss the growth of perturbations at the end of this section. It will be seen that theoretical consideration of both sources of the onset time are in qualitative agreement with the experimentally observed voltage dependence. As a first step to making this connection more concrete, we compare the $t_{onset}$ vs. $V$ scaling with expectations for the Sand's time, which corresponds to the time at which ion concentration at the surface reaches zero and the ion depleted space charge layer starts to grow. Experimentally, it is observed that at Sand's time (e.g., as estimated from the onset of dendritic growth of metal deposits) the overall system resistance increases because of low ion concentration in the space charge layer. Solving the coupled Nernst-Plank equations reveals that, $t_{sand} \sim I^{-2}$, where $I$ is the applied current.[35] While there is no analytical solution under chronoamperometry conditions, a numerical solution is possible (see supporting information **Fig. S7**), which yields the approximate scaling expression $t \sim V^{-2.3}$, which though stronger than the experimentally observed behavior is sufficiently close to what is observed to imply that the delay in onset of electroconvection reflects the time required for the space charge layer to develop near the ion-selective interface.

The mixing layer thickness and growth rate of the eddies in an electroconvective flow are important characteristics of the flow. We report experimental results in **Fig. 3b** and **3c**, respectively. Briefly, the mixing layer is here defined as the region near the interface where the most intense tracer particle motions are observed.[10] The experimental results show that incorporation of oligomer additives in liquid electrolytes reduces the thickness and growth rate of this layer —at all voltages. We also find that the while the relationship between the mixing layer growth rate and voltage is different in the electrolytes with/without polymer, the thickness increases linearly with voltage for all cases. To understand these effects, we compare with predictions from a recent theoretical analysis, which also reveal a linear scaling.[32] The analysis shows that the linear scaling arises from a balance of viscous and local electrostatic forces in non-shear electroconvective flow. An analogous scaling has been reported in previous experimental studies where the average vorticex size, or the mixing layer thickness in the case chaotic electroconvection, was reported to scale approximately linearly with applied current or voltage.[24,36] A comparison of RMS velocity distribution for control and 30% PEG300 at 3V and 8V is reported in **Fig. 3d**. The effect of voltage is twofold: firstly, the velocity near the interface increases significantly with roughly a 3-fold increase in peak velocity; secondly, the flow extends deeper in the bulk and multiple velocity peaks are observed. These features have been previously associated with multiscale and chaotic electroconvection.[24] We find that the addition of PEG300 reduces the velocity at all voltages, but unlike ultrahigh molecular weight polymers, it does not completely eliminate the flow.[22,37]

3.4 Theoretical Modeling

Finally, to understand the stabilization mechanism, we propose and investigate a theoretical model for oligomer interaction with electroconvection. Based on the experimental findings, considering only the bulk effects like viscosity increase are not sufficient to fully understand role of small MW polymers/oligomer additives. The QCM measurements suggest the presence of a thin layer with an enhanced concentration of dissolved oligomers near the surface. We therefore postulate the presence of a potential force that leads to this equilibrium layer. We model it as arising from a Van der Waals attraction between a sphere of radius $R_g$ (the radius of gyration of the oligomer) and the electrode surface. The Van der Waals potential ($\widetilde{\xi}$) is given by, $\widetilde{\xi} = \frac{-\widetilde{A}\widetilde{R}_g^3}{(\widetilde{y}+\widetilde{R}_g)^3}$, where $\widetilde{A}$ is Hamaker constant, $\widetilde{R}_g$ is radius of gyration, $k_B$ is Boltzmann constant, $T$ is temperature and $\widetilde{y}$ is the distance from the interface.[38] **Fig. 4a** shows a schematic of the stabilization mechanism. In the absence of convective flow, the van der Waals force is independent of the position x tangent to the surface and is balanced by a pressure gradient. In the presence of an electroconvective flow, the oligomer concentration isotherms are perturbed leading to a perturbation to the van der Waals force that tends to stabilize the original planar isotherms. This restoring body force opposes the convective electrolyte flow, thereby delaying electroconvection onset and increasing the system stability. Equations and parameter details are provided in the supporting information. More details about the stability analysis are provided in Mukherjee et al.[39] The effect of bulk polymer concentration on stability is reported in **Fig. 4b**. We find that maximum growth rate, which is a measure of instability in perturbation analysis, decreases as bulk polymer concentration is increased. The maximum growth rate is also an inverse measure of the time required for the onset of electroconvection. This means that at higher bulk polymer concentrations (where the maximum growth rate is lower), the time required for the onset of electroconvection should increase, something that we also observe in visualization results. Inset of **Fig. 4b** shows the dependance of this non-dimensional time with non-dimensional voltage and the corresponding scaling for control and for bulk volume fraction of $C_b$ =0.1. While the scaling varies among the two cases, -1.4 and -2.1 respectively, it is close to the experimental scaling which varies between -1.8 to -2.1 (for control and 30%PEG300 case respectively) obtained in **Fig. 3a**. Furthermore, we find that similar to our experimental results, this effect seems to saturate at higher concentrations. This is because when we increase the polymer concentration in bulk, the additional amount of polymer near the interface, which is providing stability, starts to saturate. Theoretically, this saturation is represented by excess polymer concentration per unit area near the interface and is reported in **Fig. 4c**. Details and formula for excess polymer calculation are provided in the supporting information. Experimentally, we find that the critical concentration beyond which the stabilization effects saturate is close to the overlap concentration of the polymer. **Fig. 4d** shows the specific viscosity ($\eta_{sp} = \frac{\eta_{solution}}{\eta_{solvent}} - 1$, where $\eta_{solution}$ is solution viscosity and $\eta_{solvent}$ is solvent/water viscosity) at different PEG300 concentration in electrolyte. We observe a change in slope from 1 to 1.6 between 10% to 20% PEG300 concentration, which is associated with a change from dilute to semi-dilute regime in polymer solutions.[40] This transition also aligns with the stability saturation and scaling change from $\Delta V \sim \eta^{1.8}$ to $\Delta V \sim \eta^0$ we observed in **Fig. 1c**.

## 4. Conclusions

In summary, we have investigated the effect of small molecular weight polymer additives on electroconvection at an ion selective interface. We find that oligomers additives delay the onset of electroconvection and reduce the overlimiting current. The stabilization effect initially increases with concentration but eventually saturates at higher polymer concentrations. Visualization experiments using tracer particles reveal that both applied potential and PEG concentration impact the onset and complexity of the flow. Experimentally, we obtain a scaling of electroconvection onset time with voltage as, $t_{onset} \sim V^{-\beta}$, where β lies between 1.8-2.1, indicating it is a strong decreasing function of voltage. We also find that, in addition to bulk effects of polymer additives, it is important to consider their role at the interface to fully understand the stabilization mechanism. To this end, we develop a theoretical polymer layer model to corroborate our experimental results and explore the possible physics behind the increased stability and its eventual saturation. Through perturbation analysis on this model, we find that the presence of a polymer layer at the interface adds a restoring body force to the stokes equation which opposes the convective electrolyte flow, thereby delaying electroconvection onset and increasing the system stability.

**Supporting Information**
Current Voltage analysis (Fig S1-S4); Submitted Visualization Video details (cc 1v; 30% 8v); GEIS Modeling (Fig S5); Electrolyte Viscosity and Conductivity(Table S1 and Table S2); QCM-D Modeling (Fig S6, Table S3 and S4); Sands Time Simulation (Fig S7); Electrolyte Rheology (Fig S8); Theoretical Modeling (Table S5 and Fig S9)

19. Maletzki, F., Rösler, H. W. & Staude, E. Ion transfer across electrodialysis membranes in the overlimiting current range: stationary voltage current characteristics and current noise power spectra under different conditions of free convection. *J. Memb. Sci.* **71**, 105–116 (1992).
20. Khurana, R., Schaefer, J. L., Archer, L. A. & Coates, G. W. Suppression of lithium dendrite growth using cross-linked polyethylene/poly(ethylene oxide) electrolytes: A new approach for practical lithium-metal polymer batteries. *J. Am. Chem. Soc.* **136**, 7395–7402 (2014).
21. Wei, S. *et al.* Stabilizing electrochemical interfaces in viscoelastic liquid electrolytes. *Sci. Adv.* **4**, 1–9 (2018).
22. Warren, A., Zhang, D., Choudhury, S. & Archer, L. A. Electrokinetics in Viscoelastic Liquid Electrolytes above the Diffusion Limit. *Macromolecules* (2019). doi:10.1021/acs.macromol.9b00536
23. Huth, J. M., Swinney, H. L., McCormick, W. D., Kuhn, A. & Argoul, F. Role of convection in thin-layer electrodeposition. *Phys. Rev. E* **51**, 3444–3458 (1995).
24. Warren, A., Sharma, A., Zhang, D., Li, G. & Archer, L. A. Structure and Dynamics of Electric-Field-Driven Convective Flows at the Interface between Liquid Electrolytes and Ion-Selective Membranes. *Langmuir* **37**, 5895–5901 (2021).
25. Thielicke, W. & Stamhuis, E. J. PIVlab – Towards User-friendly, Affordable and Accurate Digital Particle Image Velocimetry in MATLAB. *J. Open Res. Softw.* **2**, (2014).
26. Rodahl, M. & Kasemo, B. A simple setup to simultaneously measure the resonant frequency and the absolute dissipation factor of a quartz crystal microbalance. *Rev. Sci. Instrum.* **67**, 3238–3241 (1996).
27. Voinova, M. V, Rodahl, M., Jonson, M. & Kasemo, B. Viscoelastic Acoustic Response of Layered Polymer Films at Fluid-Solid Interfaces: Continuum Mechanics Approach. *Phys. Scr.* **59**, 391–396 (1999).
28. Rubinstein, I. & Zaltzman, B. Electro-osmotically induced convection at a permselective membrane. *Phys. Rev. E - Stat. Physics, Plasmas, Fluids, Relat. Interdiscip. Top.* **62**, 2238–2251 (2000).
29. Li, G. *et al.* Stabilizing electrochemical interfaces in viscoelastic liquid electrolytes. *Sci. Adv.* **4**, eaao6243 (2018).
30. Likhtman, A. E. & Graham, R. S. Simple constitutive equation for linear polymer melts derived from molecular theory: Rolie-Poly equation. *J. Nonnewton. Fluid Mech.* **114**, 1–12 (2003).
31. Dutta, A. K., Nayak, A. & Belfort, G. Viscoelastic properties of adsorbed and cross-linked polypeptide and protein layers at a solid-liquid interface. *J. Colloid Interface Sci.* **324**, 55–60 (2008).
32. Weber, N., Wendel, H. P. & Kohn, J. Formation of viscoelastic protein layers on polymeric surfaces relevant to platelet adhesion. *J. Biomed. Mater. Res. - Part A* **72**, 420–427 (2005).
33. Malmström, J., Agheli, H., Kingshott, P. & Sutherland, D. S. Viscoelastic modeling of highly hydrated laminin layers at homogeneous and nanostructured surfaces: Quantification of protein layer properties using QCM-D and SPR. *Langmuir* **23**, 9760–9768 (2007).
34. Vidyasagar, A., Sung, C., Losensky, K. & Lutkenhaus, J. L. PH-dependent thermal transitions in hydrated layer-by-layer assemblies containing weak polyelectrolytes.

**Figure File**

# Electroconvective flow in presence of polyethylene glycol oligomer additives

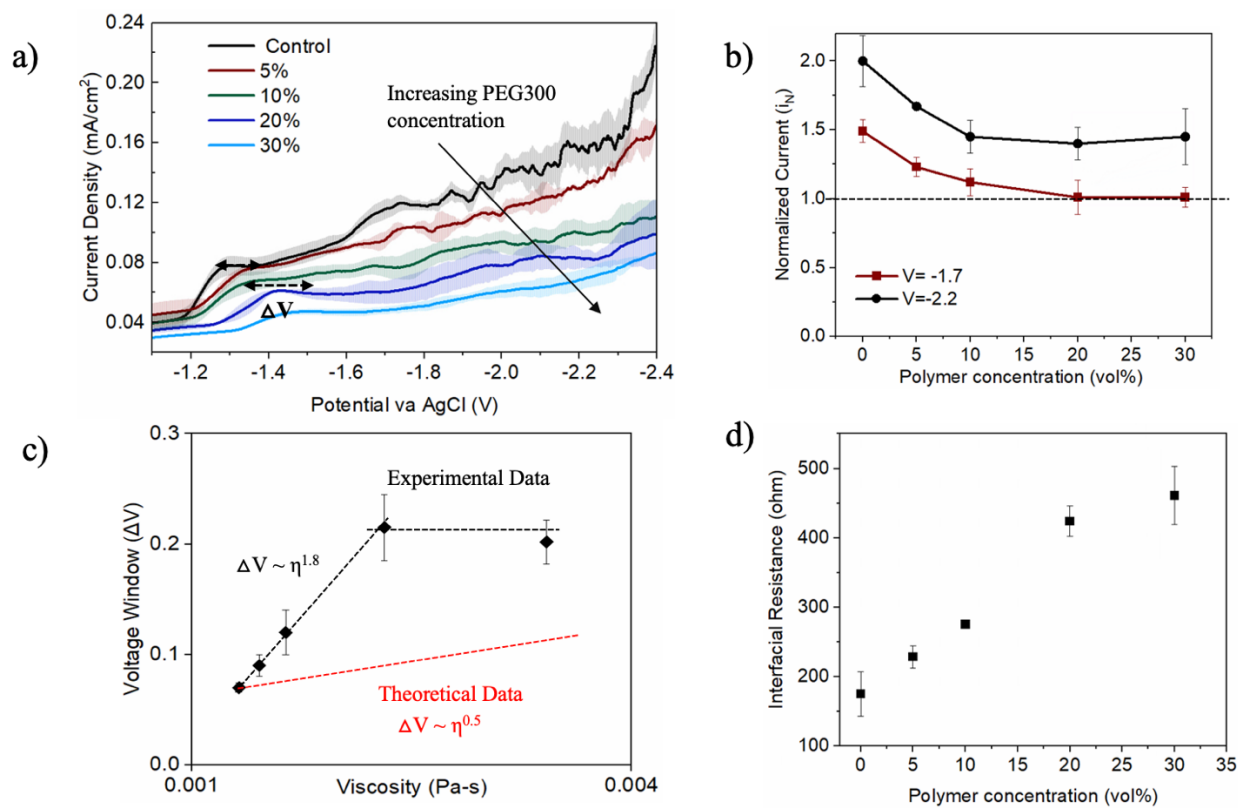

Figure 1: a)- Current voltage response for 1mM $ZnSO_4$ with different concentrations of PEG300 additive in the electrolyte. b)- Normalized current at different voltages in the over-limiting region. $i_N = 1$ (dashed line) indicates no electroconvection. c)- Voltage window extension vs viscosity scaling obtained experimentally (black), and from theoretical prediction (red) for Newtonian electrolytes. d)- Interfacial resistance as a function of polymer concentration obtained from EIS experiments.

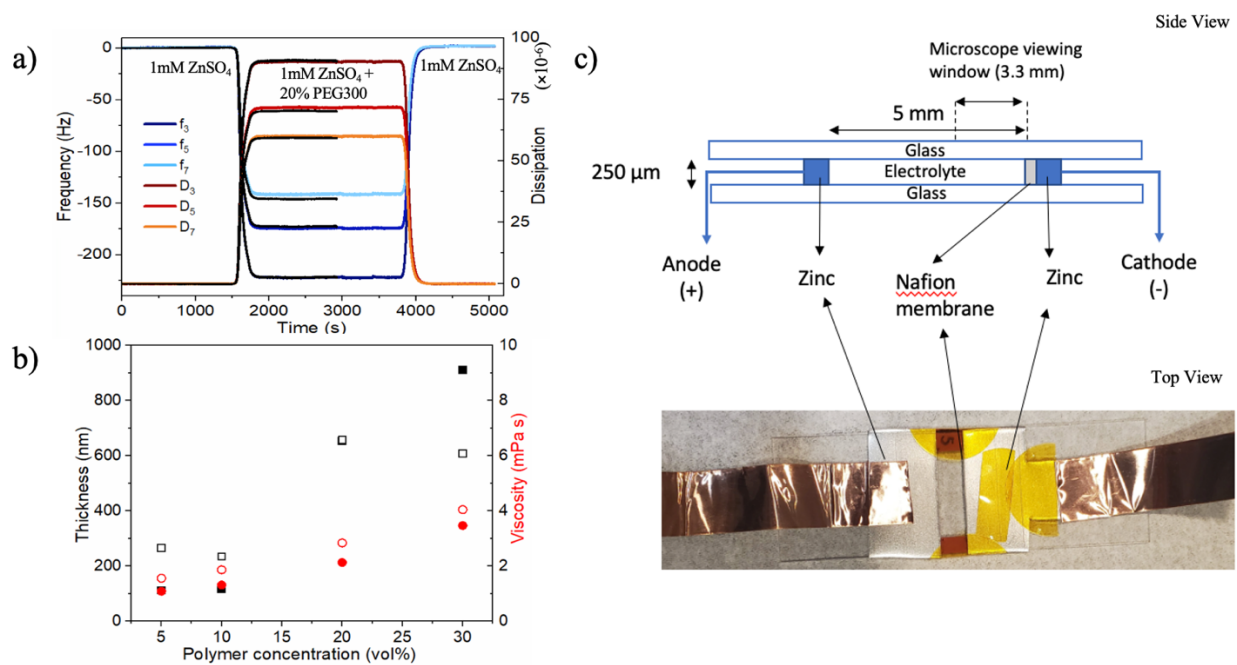

Figure 2: a)- Frequency and dissipation shift for third, fifth and seventh harmonics on flowing 1mM ZnSO$_4$+20% PEG300 solution after salt baseline. Data obtained from QCM-D experiments. b)- Layer viscosity and thickness as a function of polymer concentration obtained from Voigt single layer model - filled symbols show data obtained from inbuilt Voigt model in Qtools software; open symbols show data obtained by solving Voigt equations in Matlab and considering small layer elasticity. c)- Visualization cell schematic and actual device used to record flow videos.

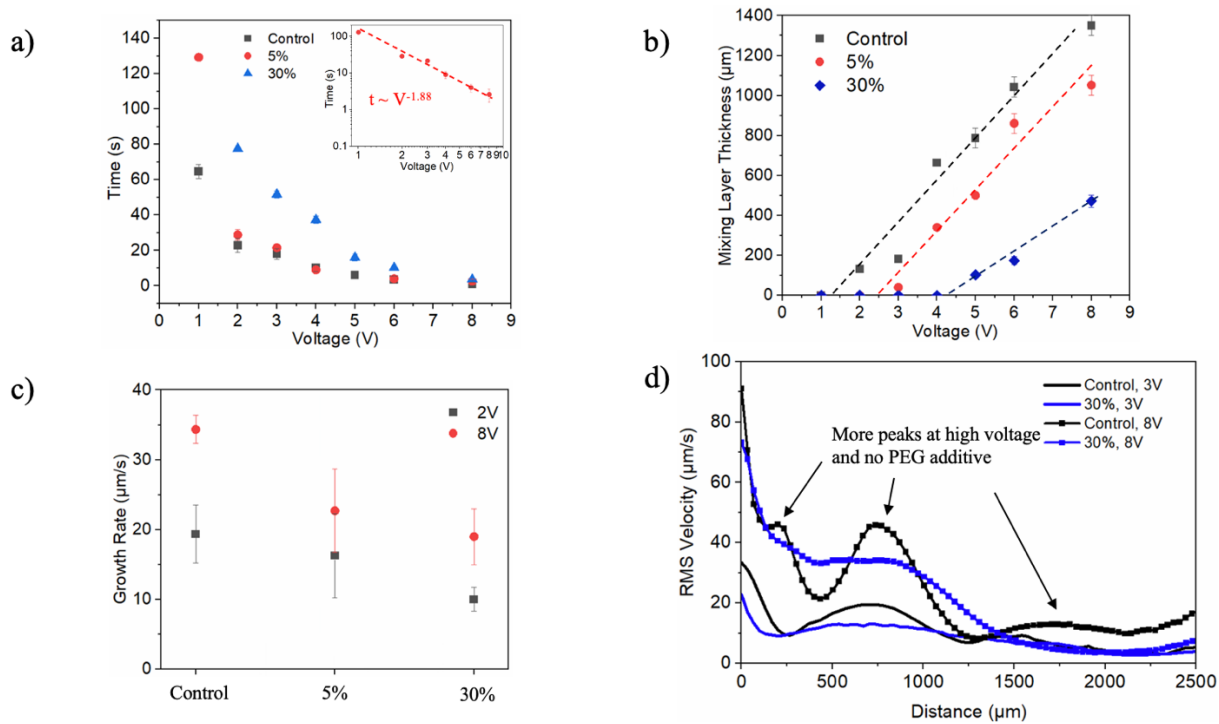

Figure 3: a)- Electroconvective flow onset time as a function of voltage obtained from visualization experiments. Inset: Log-log plot for time vs voltage for 1mM $ZnSO_4$ +5% PEG300 data and the corresponding power law scaling. b)- Mixing layer thickness 20 seconds after connecting to the power supply. Dashed lines are added for guidance and do not represent data fitting. c)- Average growth rate of mixing layer at different voltages. d)- RMS velocity distribution for 1mM $ZnSO_4$ (control) and 1mM $ZnSO_4$ +30% PEG300 (30%) electrolyte at 3 and 8V.

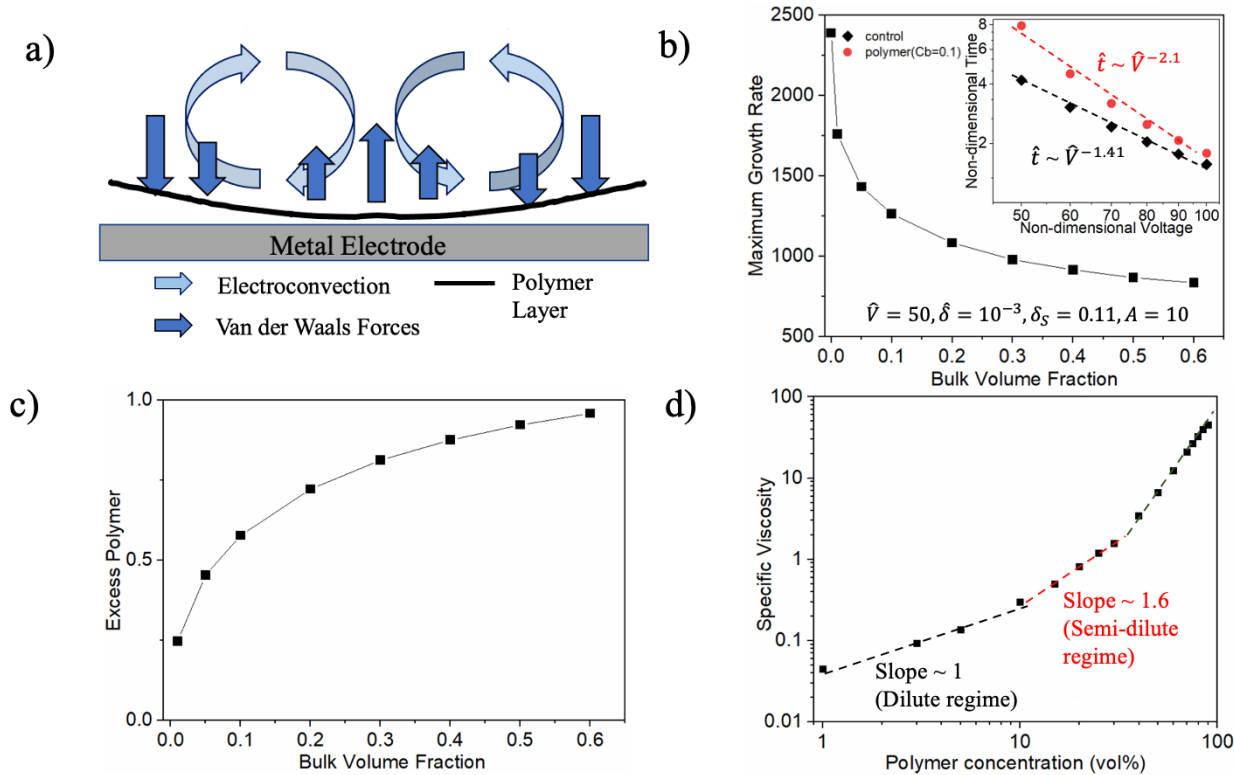

Figure 4: a)- Direction of electroconvective flow and Van der Waals force in the theoretical model. Van der Waals forces of attraction between the metal electrode and the polymer layer at the interface result in a stabilizing volumetric force on the electrolyte. This force acts as a restoring force to maintain the polymer layer thickness, thereby suppresses electroconvection and increasing system stability. b)- Maximum growth rate obtained from perturbation analysis. Inset: Log-log plot of non-dimensional time vs non-dimensional voltage and the corresponding power law scaling for control and bulk volume fraction, $C_b$ = 0.1, case. c)- Excess polymer near the interface as a function of bulk polymer concentration. Theoretical results (4b and 4c) were obtained using non-dimensional parameters: voltage $(\hat{V})$ = 50, double layer thickness $(\hat{\delta})$ = $10^{-3}$ , space charge thickness $(\hat{\delta}_S)$ = 0.11 and Hamaker constant $(A)$ = 10. d)- Specific viscosity data obtained from rheological experiments shows a change from dilute to semi-dilute regime as polymer concentration increases.

**Supporting information:**

1. Current-Voltage curve analysis – Normalized Current density ($i_N$) showing limiting plateau or voltage window measurement ($\Delta V$). All the i-V curves were normalized by dividing the current by limiting current for each case. Length of the limiting plateau as a voltage window is then calculated as $V_2-V_1$ where, $V_1$ and $V_2$ are the points where $i_N = 1$ intersects with the normalized i-V curve.

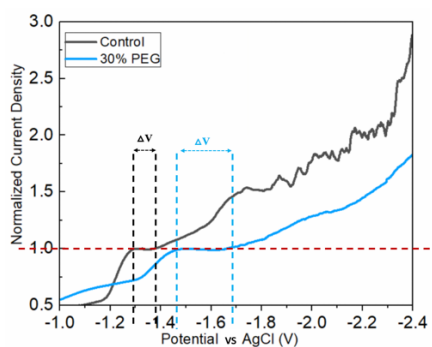

Fig. S1 – Voltage window ($\Delta V$) measurement for control (1mM $ZnSO_4$ at 1V); 30%PEG (1mM $ZnSO_4$ + 30% PEG300).

2. I-V data for 10mM $ZnSO_4$ – A similar i-V response was observed for higher salt concentrations. Fig. S2 shows the i-v data for 10mM $ZnSO_4$ at different PEG300 concentrations at a scan rate of 1mV/s. An extension in limiting plateau and decrease in overlimiting current is observed in the presence of PEG300.

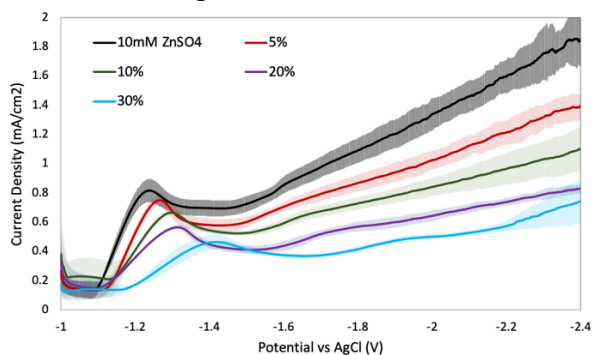

Fig. S2 – I-v data for 10mM $ZnSO_4$ at different PEG 300 additive concentrations.

3. No Nafion membrane results – To show the importance of Nafion membrane, we conducted a control case without nafion for 10mM $ZnSO_4$ as shown in Fig. S3. As seen below, we still see the extension in the limiting plateau and decrease in overliming current in the presence of PEG300. However, the overlimiting current increases significantly because of aggressive dendrite growth as seen in Fig. S4.

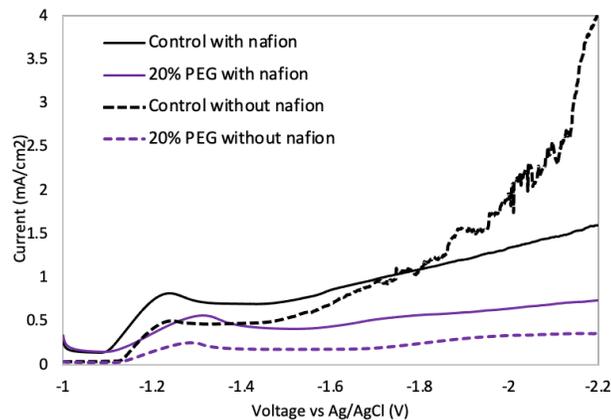

Fig S3 – Control (10mM ZnSO$_4$) with and without PEG and Nafion membrane.

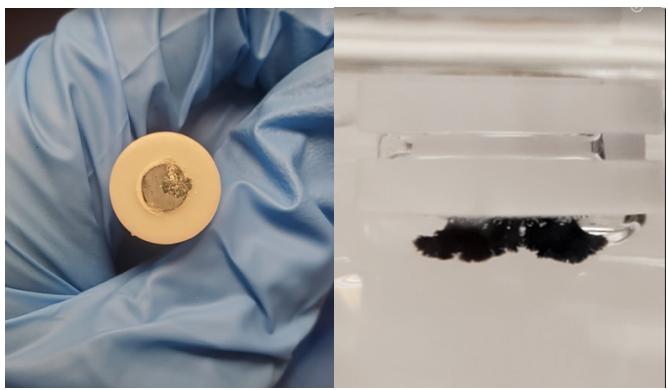

Fig S4- Significant dendrite growth on the electrode after i-V run for 10mM ZnSO$_4$. This issue can be prevented by using Nafion membrane.

4. Visualization experiment videos – Three videos are submitted with the supporting information spanning different voltages and additive concentration. Tittle format – cc 1v (control case, 1mM ZnSO$_4$ at 1V); 30% 8v (1mM ZnSO$_4$ + 30% PEG300 at 8V). Speed enhanced four times (to 100fps) compared to the actual experimental rate (25 fps). The visualization videos are processed using the PIVlab software in Matlab.

5. GEIS modeling – Nyquist impedance plots at different polymer concentrations are reported in Fig. S5. An equivalent circuit model - $R1 + R2/Q2 + R3/Q3$ was used to fit the data, where R is the resistance and Q is the non-ideal capacitance. The circuit model was chosen based on previous studies at zinc interfaces and is guided by possible surface resistances that can be present at a zinc metal interface.[1–3] The 3 components considered are bulk resistance (R1), SEI layer (containing ZnO, Zn(OH)$_2$ etc.) at the surface (R2/Q2) and charge transfer/ interfacial resistance (R3/Q3) respectively. For all 5 concentrations, a good fit was obtained for α value between 1 to 0.85, where α=1 denotes an ideal capacitance.

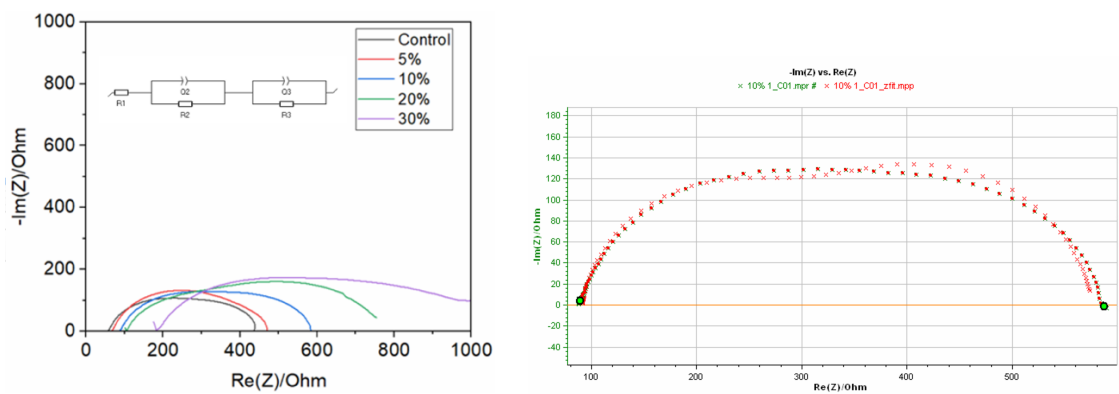

Fig S5 – a)- Nyquist impedance at different polymer concentrations. b)- Experimental and fitted data for 1mM $ZnSO_4$+10% PEG300.

6. Rheology data – Bulk viscosity of electrolytes was measured using a parallel plate geometry (diameter 50mm, gap 500 μm) on DHR3 rheometer from TA instruments at a shear rate of 300/s. Table S1 shows the viscosity as a function of polymer concentration.

Table S1

| Concentration of PEG300 in 1mM $ZnSO_4$ | Viscosity (mPa-s) |
| --- | --- |
| 0% (1mM znso4 without PEG) | 1.3 |
| 1% | 1.337 |
| 3% | 1.398 |
| 5% | 1.452 |
| 10% | 1.658 |
| 15% | 1.914 |
| 20% | 2.312 |
| 25% | 2.792 |
| 30% | 3.281 |
| 40% | 5.64 |
| 50% | 9.75 |
| 60% | 16.93 |
| 70% | 27.88 |
| 75% | 35.1 |
| 80% | 42.47 |
| 85% | 52 |
| 90% | 59.12 |

7. Conductivity measurements: Liquid electrolyte conductivity at 20 C.

Table S2

| Concentration of PEG300 in 1mM ZnSO$_4$ | Conductivity (mS/cm) |
|---|---|
| 0% (1mM znso4 without PEG) | 0.203243 |
| 5% | 0.188775 |
| 10% | 0.172614 |
| 20% | 0.09750295 |
| 30% | 0.0690223 |

8. Ellipsometry- Dry polymer layer thickness was measured by using Alpha SE Ellipsometer. A silicon wafer with an area of 4cm$^2$ was used. For each silicon wafer, we tested 6 points and get the average thickness. For each kind of concentration, we tested 3 samples, which means totally we tested 18 times for the PEG300 adsorption thickness and then extracted the average value. These measurements were done in the absence of Zinc sulfate salt.

Table S3

| PEG 300 Concentration | Dry layer thickness from Ellipsometry (nm) | Layer thickness from inbuilt Voigt model from QCM-D (nm) |
|---|---|---|
| 5% | 0.74 ±0.2 | 110 |
| 10% | 1.71 ±0.3 | 118 |
| 20% | 4.06 ±0.3 | 659 |
| 30% | 6.33 ±0.35 | 911 |

9. QCM- D Modeling –

i)- Q-tools inbuilt Voigt model: Fig S6 shows the raw and fitted data for 1mM ZnSO$_4$+20% PEG300 electrolyte obtained from Q-tools. Black lines represent the fitted values for 3$^{rd}$, 5$^{th}$ and 7$^{th}$ harmonics.

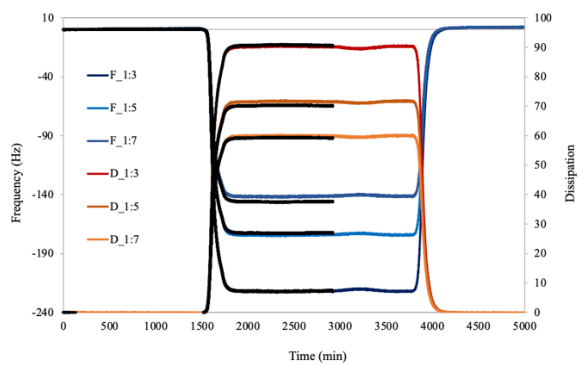

Fig S6 – Experimental and fitted values for 1mM ZnSO$_4$+20% PEG300 for different harmonics obtained from inbuilt Voigt model in Qtools.

ii)- Voigt model solution from Matlab: The coupled Voigt model equations shown below were solved using Matlab.[4] The modeled parameters – thickness, viscosity and elasticity are reported in Table S4.

$$f = nf_0$$

$$\Delta f = \frac{Im(\beta_q)}{2\pi d_q \rho_q}$$

$$\Delta D = -\frac{Re(\beta_q)}{\pi f d_q \rho_q}$$

$$\beta_q = \frac{(2\pi f \eta \xi_1 - i\mu\xi_1)(1 - \alpha_q \exp(2\xi_1 d))}{2\pi f(1 + \alpha_q \exp(2\xi_1 d))} - \beta_0$$

$$\alpha_q = \frac{2\pi f \eta \xi_1 - i\mu\xi_1 + 2\pi f \eta_l \xi_2}{2\pi f \eta \xi_1 - i\mu\xi_1 - 2\pi f \eta_l \xi_2}$$

$$\xi_0 = \sqrt{\frac{i 2\pi f \rho_0}{\eta_0}}$$

$$\beta_0 = -\eta_0 \xi_0$$

$$\xi_1 = \sqrt{-\frac{(2\pi f)^2 \rho}{\mu + i 2\pi f \eta}}$$

$$\xi_2 = \sqrt{\frac{i 2\pi f \rho_l}{\eta_l}}$$

Here, $f_0$ is the frequency of oscillation of the quartz crystal, $n$ refers to the harmonic number. $\Delta f$ and $\Delta D$ are the changes in frequency when water flowing over the Quartz crystal is replaced with PEG solution. $d_q$ and $\rho_q$ are the thickness and density of the Quartz crystal. $\eta$, $\mu$, $\rho$ and $d$ are the viscosity, elasticity, density and thickness of the polymer layer formed at the surface of the Quartz crystal. $\eta_l$ and $\rho_l$ are the viscosity and density of the bulk fluid, while $\eta_0$ and $\rho_0$ are the viscosity and density of water that flows over the Quartz crystal initially. $\beta_0$ is a term due to the initial shear force when water flows over the Quartz crystal. The following expression is minimized using the fminsearch function in MATLAB. The function solves for the values of layer thickness, volume fraction and elasticity that minimize F.

$$F = \sum_{i=3,5,7}\left[\left(\frac{\Delta f_{experimental,i} - \Delta f_{calculated,i}}{\Delta f_{experimental,i}}\right)^2 + \left(\frac{\Delta D_{experimental,i} - \Delta D_{calculated,i}}{\Delta D_{experimental,i}}\right)^2\right]$$

The layer density and viscosity are solved using the following equations:

$$\rho = \rho_0 + (\rho_{PEG} - \rho_0)\phi$$

$$\frac{\eta - \eta_0}{\eta_l - \eta_0} = \frac{\phi}{\phi_l}$$

Here, $\phi$ is the volume fraction of PEG in the polymer layer, while $\phi_l$ is the volume fraction of PEG in the bulk PEG solution. $\rho_{PEG}$ is the density of pure PEG solution.

Table S4

| PEG300 Concentration | Matlab Voigt Model Solution – Thickness (nm) | Matlab Voigt Model Solution – Viscosity (mPa s) | Matlab Voigt Model Solution – Elasticity (Pa) |
|---|---|---|---|
| 5% | 266 | 1.552 | 0.006788 |
| 10% | 235 | 1.861 | 0.00234 |
| 20% | 657 | 2.834 | 0.164972 |
| 30% | 603 | 4.045 | 0.005985 |

10. Theoretical electroconvection onset time: Sand's time under potentiostatic conditions was obtained by numerically solving the following equations and boundary conditions for concentration. The sand's time is defined as the time at which concentration at the surface becomes zero. Fig S7 shows the scaling obtained.

$$\frac{\partial c}{\partial t} = D \frac{\partial^2 c}{\partial y^2}$$

$$\frac{\partial c}{\partial y}\bigg|_{y=0} \int_0^1 \frac{1}{c} dy = \hat{V}$$

BCs: $c|_{y=1/2} = 1$ ; $c(t=0) = 1$

Here, c is the salt concentration, t is the time, L is interelectrode distance, $\hat{V}$ is the applied potential. The variables are normalized as follows: concentration by the initial uniform concentration c0, voltage by the thermal voltage RT/F, length by L, and time by $L^2/D$ and D is the diffusion constant. In comparison, the dimensionless Sand's time for a galvanostatic polarization at a limiting current is $\tau = \pi/16$.

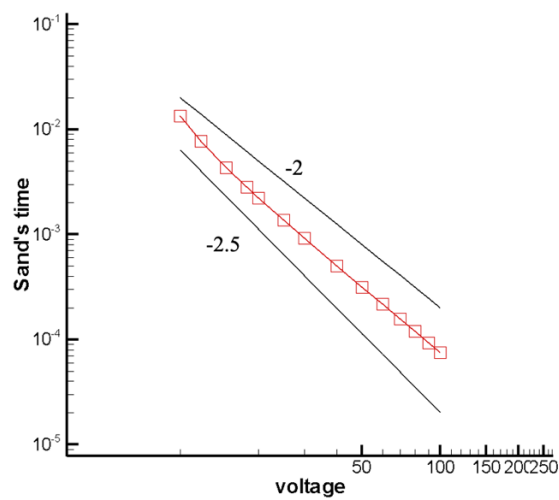

Fig S7 – Sand's time under potentiostatic conditions vs voltage.

11. Rheology – Strain sweep and frequency sweep were conducted on ARES rheometer from TA instruments using 50mm cone and plate geometry. We find that even for the highest oligomer concentration tested, 1mM ZnSO$_4$ + 30% PEG300, the sample is predominately viscous with very small elastic contribution.

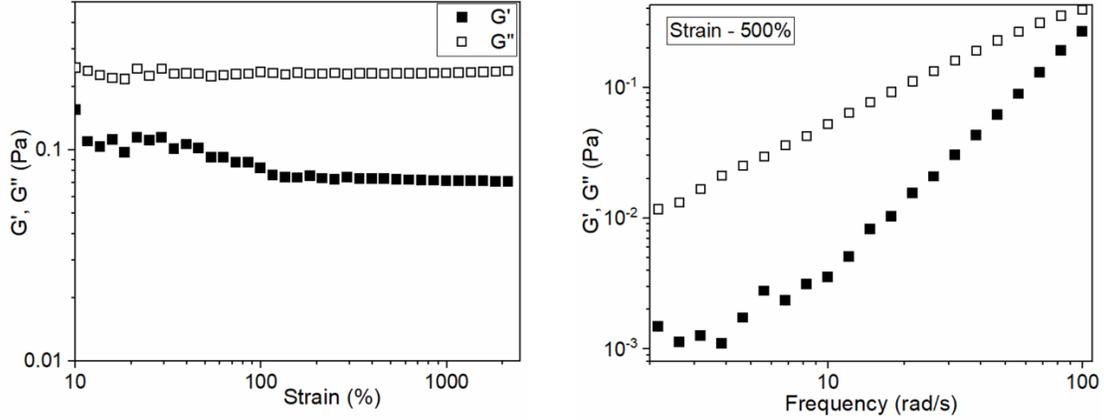

Fig S8 – Strain sweep and frequency sweep for 1mM ZnSO$_4$ + 30% PEG300.

12. Theoretical Modeling: Linear stability analysis of the system with Van der Waals forces at the ion-selective surface: The base state polymer concentration is obtained by setting the net flux to 0 and solving the resulting equations:

$$D(\phi)\frac{d\Psi_b}{dy} + \Psi_b \frac{d\xi}{dy} = 0,$$
$$D(\phi) = \frac{d(\phi Z)}{d\phi},$$
$$Z = \frac{\pi}{nk_BT} = 1 + 0.5\phi^{\frac{5}{4}}$$

Here, the Van der Waal's potential is $\xi \left(= \frac{-AR_g^3}{(y+R_g)^3}\right)$. A is the non-dimensional Hamaker constant, $R_g$ is the radius of gyration of the polymer, $y$ is the normal distance from the interface, $k_B$ is the Boltzmann constant and $n$ is the number of polymer molecules per unit volume. $D(\phi)$ is the diffusion coefficient as defined in Russell (1991), $\phi$ is the volume fraction of the polymer, $\Psi_b$ is the base state polymer concentration and $Z$ is the compressibility factor.[5] An equation of state for PEG is used to obtain the expression for the compressibility factor.[6,7]

Modified Stokes equation:

$$-\nabla p + \nabla^2 \vec{u} + \nabla \hat{V} \nabla^2 \hat{V} - \beta \Psi \nabla \xi = 0$$

Here, β is a non-dimensional value defined in terms of viscosity and scalings for potential, concentration, length and velocity. $\beta = \left(\frac{\Psi_{sc}\xi_{sc}L}{\eta u_{sc}}\right)$. Ψ is the perturbed polymer concentration. $\Psi_{sc}$ is the scaling for polymer concentration. $\xi_{sc} = k_BT$ is the scaling for Van der Waals potential. $\eta$ is the viscosity of the electrolyte. The van der Waals force acting on each sphere is multiplied by the concentration of the polymer spheres to obtain a continuum representation of this force. While the van der Waals force $(-\beta\Psi\nabla\xi)$ is a part of the equation which is solved in the entire electrolyte, its value is 0 in the bulk electrolyte as the van der Waals force is a short-range force that rapidly decays away from the metal electrode at y=0 (as the van der Waals potential decays as $1/y^3$). The polymer concentration decays rapidly away from the electrode surface, but it is not a delta function. An inset has been added to fig S5 to show the variation near the electrode surface. The volume fraction shown in these figures is the volume fraction of a sphere that encircles a single polymer. It has a radius of gyration R g and can also be described as the ratio of the concentration to the overlap concentration of the polymer. For more concentrated solutions that exist near the electrode, these spheres can overlap, hence the volume fraction can be greater than 1.

Table S5
(Dimensional parameters are represented using a ~ symbol over the variable.)

| Non-dimensional parameters | Expressions in terms of dimensional values |
| --- | --- |
| $\hat{V}$ | $\hat{V} = \dfrac{\tilde{V}}{RT/F}$ |
| $A$ | $A = \dfrac{\tilde{A}}{k_B T}$ |
| $\hat{\delta}$ | $\hat{\delta} = \dfrac{\tilde{\delta}}{L} = \dfrac{\sqrt{\dfrac{\epsilon_r \epsilon_0 RT}{2F^2 C_0}}}{L}$ |
| $y, \delta_s, R_g$ | $y = \dfrac{\tilde{y}}{L}, \delta_s = \dfrac{\tilde{\delta_s}}{L}, R_g = \dfrac{\widetilde{R_g}}{L}$ |
| $\Psi$ | $\Psi = \dfrac{\tilde{\Psi}}{\Psi_{sc}}$ |
| $\xi$ | $\xi = \dfrac{\tilde{\xi}}{k_B T}$ |
| $\vec{u}$ | $\vec{u} = \dfrac{\tilde{\vec{u}}}{u_{sc}} = \dfrac{\tilde{\vec{u}}}{\left(\epsilon_r \epsilon_0 (RT)^2 / F^2 \eta L\right)}$ |
| $p$ | $p = \dfrac{\tilde{p}}{(\eta u_{sc}/L)}$ |

$\varepsilon_r$ is the dielectric constant, $\varepsilon_0$ is the vacuum permittivity, R is the universal gas constant, T is the absolute temperature, F is Faraday's constant, L is the inter-electrode distance and $\eta$ is the fluid viscosity. The value of A ($= \dfrac{\tilde{A}}{k_B T}$) is taken to be 1 which is a typical value for the Hamaker constant. The value of A was calculated in Mukherjee et al. 2022 and was found to be 2. A=1 and A=10 can represent slightly stronger and weaker attractive forces. The non-dimensional value of the applied voltage $\hat{V}$ is 50, which corresponds to a dimensional value of 1.3V. The non-dimensional double layer thickness and the non-dimensional radius of gyration are both taken to be 0.001, which corresponds to a dimensional value of 1 μm. These are slightly larger than typical values, however, they are significantly smaller than the inter-electrode distance, which ensures that we can describe the physics of the problem accurately, while also ensuring that the numerical problem is solvable.

The Poisson and Nernst Planck equations are used to solve the electric potential and ion concentration. The continuity and modified Stokes equations are used to solve for velocity and pressure.

$$\frac{\partial C_C}{\partial t} + Pe\vec{u}.\vec{\nabla} C_C = \frac{D+1}{2}\vec{\nabla}.[\vec{\nabla}C_C + C_C\vec{\nabla}\hat{V}]$$

$$\frac{\partial C_A}{\partial t} + Pe\vec{u}.\vec{\nabla} C_A = \frac{D+1}{2D}\vec{\nabla}.[\vec{\nabla}C_A - C_A\vec{\nabla}\hat{V}]$$

$$2\hat{\delta}^2 \nabla^2 \hat{V} = C_A - C_C$$

$$\vec{\nabla}.\vec{u} = 0$$

$C_C$ and $C_A$ are the concentration of the cation and anion. Pe is the Peclet number, and its value is approximately 0.5 for salt solutions. D is the ratio of cation and anion diffusivities, and its value is taken to be 1. The cation concentration is set at a fixed value (1) at the ion-selective surface and the anion flux is 0. The potential is 0. At the counter electrode, ion concentrations are set to 1 and the potential is set to the value of the applied voltage.

Perturbations of the form $\zeta = \widehat{\zeta(y)} \exp(\sigma t + ikx)$ are used for cation and anion concentration, electric potential, velocity, pressure and polymer concentration. These perturbations are substituted into the governing equations. The resulting equations are used to solve for the growth rate ($\sigma$) as a function of the wavenumber($k$). The highest value of $\sigma$ represents the most unstable mode.

Further details about the linear stability analysis can be found in Li et al. (2021), Li et al. (2022) and Mukherjee et al. (2022).[8–10]

13. Excess polymer in theoretical analysis: Excess polymer concentration per unit area of the electrode surface is calculated as $\int(C - C_b)dy$, where $C$ is the concentration profile with a build-up of polymers near the electrode and $C_b$ is the bulk concentration (which is also the initial concentration). Integration is carried out over the whole domain from y = 0 to y = 1. Fig S9 shows the volume fraction profile for bulk volume fraction = 0.6.

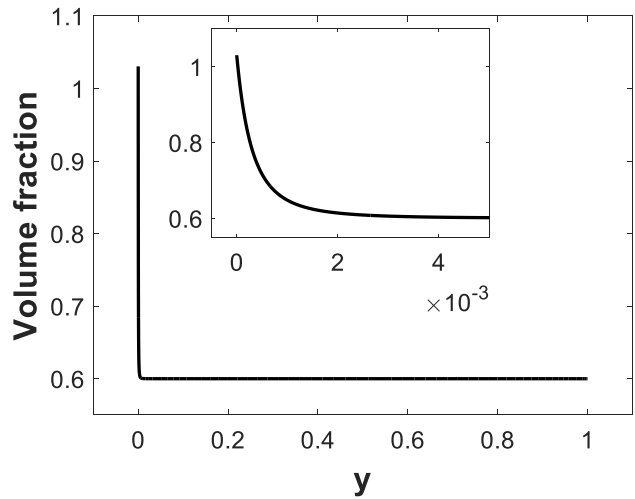

Fig S9 – (a) Polymer volume fraction profile for bulk concentration = 0.6 and Hamaker constant, A =1. (b) Same plot showing the increase in the concentration in the space charge layer. The volume fraction is the ratio of the polymer concentration to the overlap concentration, so its value can be greater than 1 when the polymer concentration is very high (such as near the electrode).